\documentstyle[12pt]{article}
\newcommand{\nc}{\newcommand}
\nc{\renc}{\renewcommand}

\nc{\half}{{\textstyle{1\over2}}}

\nc{\etal}{\mbox{\it et al. }}
\nc{\ie}{{\it i.e.}}
\nc{\eg}{{\it e.g.}}

\renc{\thefootnote}{\arabic{footnote}}
\nc{\capt}[1]{{\bf Figure.} {\small\sl #1}}

\nc{\eqs}[2]{\mbox{Eqs.~(\ref{#1},\,\ref{#2})}}
\nc{\eq}[1]{\mbox{Eq.~(\ref{#1})}}

\nc{\figs}[2]{\mbox{Figs.~(\ref{#1},\,\ref{#2})}}
\nc{\fig}[1]{\mbox{Fig~.(\ref{#1})}}

\nc{\tag}[1]{\label{#1} \marginpar{{\footnotesize #1}}}
\nc{\mtag}[1]{\label{#1} \mbox{\marginpar{{\footnotesize #1}}}}
\renc{\baselinestretch}{1.2}
\newlength{\overeqskip}
\newlength{\undereqskip}
\nc{\be}[1]{\begin{equation} \mbox{$\label{#1}$}}
\nc{\bea}[1]{\begin{eqnarray} \mbox{$\label{#1}$}}
\nc{\Section}[2]{\section{#2}\label{#1}}
\nc{\Bibitem}[1]{\bibitem{#1}}
\nc{\Label}[1]{\label{#1}}

\nc{\eea}{\vspace{\undereqskip}\end{eqnarray}}
\nc{\ee}{\vspace{\undereqskip}\end{equation}}
\nc{\bdm}{\begin{displaymath}}
\nc{\edm}{\end{displaymath}}
\nc{\dpsty}{\displaystyle}
\nc{\bc}{\begin{center}}
\nc{\ec}{\end{center}}
\nc{\ba}{\begin{array}}
\nc{\ea}{\end{array}}
\nc{\bab}{\begin{abstract}}
\nc{\eab}{\end{abstract}}
\nc{\btab}{\begin{tabular}}
\nc{\etab}{\end{tabular}}
\nc{\bit}{\begin{itemize}}
\nc{\eit}{\end{itemize}}
\nc{\ben}{\begin{enumerate}}
\nc{\een}{\end{enumerate}}
\nc{\bfig}{\begin{figure}}
\nc{\efig}{\end{figure}}
\nc{\arreq}{&\!=\!&}
\nc{\arrmi}{&\!-\!&}
\nc{\arrpl}{&\!+\!&}
\nc{\arrap}{&\!\!\!\approx\!\!\!&}
\nc{\non}{\nonumber\\*}
\nc{\align}{\!\!\!\!\!\!\!\!&&}

\def\lsim{\; \raise0.3ex\hbox{$<$\kern-0.75em
      \raise-1.1ex\hbox{$\sim$}}\; }
\def\gsim{\; \raise0.3ex\hbox{$>$\kern-0.75em
      \raise-1.1ex\hbox{$\sim$}}\; }
\nc{\DOT}{\hspace{-0.08in}{\bf .}\hspace{0.1in}}
\nc{\Laada}{\hbox {$\sqcap$ \kern -1em $\sqcup$}}
\nc\loota{{\scriptstyle\sqcap\kern-0.55em\hbox{$\scriptstyle\sqcup$}}}
\nc\Loota{{\sqcap\kern-0.65em\hbox{$\sqcup$}}}
\nc\laada{\Loota}
\nc{\qed}{\hskip 3em \hbox{\BOX} \vskip 2ex}

\nc{\real}{{\rm I \! R}}
\nc{\Z}{{\sf Z \!\!\! Z}}
\nc{\complex}{{\rm C\!\!\! {\sf I}\,\,}}
\def\bigid{\leavevmode\hbox{\small1\kern-3.8pt\normalsize1}}
\def\id{\leavevmode\hbox{\small1\kern-3.3pt\normalsize1}}
\nc{\slask}{\!\!\!/}
\nc{\bis}{{\prime\prime}}
\nc{\pa}{\partial}
\nc{\na}{\nabla}
\nc{\ra}{\rangle}
\nc{\la}{\langle}
\nc{\goto}{\rightarrow}
\nc{\swap}{\leftrightarrow}

\nc{\EE}[1]{ \mbox{$\cdot10^{#1}$} }
\nc{\abs}[1]{\left|#1\right|}
\nc{\at}[2]{\left.#1\right|_{#2}}
\nc{\norm}[1]{\|#1\|}
\nc{\abscut}[2]{\Abs{#1}_{\scriptscriptstyle#2}}
\nc{\vek}[1]{{\rm\bf #1}}
\nc{\integral}[2]{\int\limits_{#1}^{#2}}
\nc{\inv}[1]{\frac{1}{#1}}
\nc{\dd}[2]{{{\partial #1}\over{\partial #2}}}
\nc{\ddd}[2]{{{{\partial}^2 #1}\over{\partial {#2}^2}}}
\nc{\dddd}[3]{{{{\partial}^2 #1}\over
        {\partial #2 \partial #3}}}
\nc{\dder}[2]{{{d #1}\over{d #2}}}
\nc{\ddder}[2]{{{d^2 #1}\over{d {#2}^2}}}
\nc{\dddder}[3]{{d^2 #1}\over
        {d #2 d #3}}
\nc{\dx}[1]{d\,^{#1}x}
\nc{\dy}[1]{d\,^{#1}y}
\nc{\dz}[1]{d\,^{#1}z}
\nc{\dl}[1]{\frac{d\,^{#1}l}{(2\pi)^{#1}}}
\nc{\dk}[1]{\frac{d\,^{#1}k}{(2\pi)^{#1}}}
\nc{\dq}[1]{\frac{d\,^{#1}q}{(2\pi)^{#1}}}

\nc{\cc}{\mbox{$c.c.$ }}
\nc{\hc}{\mbox{$h.c.$ }}
\nc{\cf}{cf.\ }
\nc{\erfc}{{\rm erfc}}
\nc{\Tr}{{\rm Tr\,}}
\nc{\tr}{{\rm tr\,}}
\nc{\pol}{{\rm pol}}
\nc{\sign}{{\rm sign}}
\nc{\bfT}{{\bf T }}

\nc{\cA}{{\cal A}}
\nc{\cB}{{\cal B}}
\nc{\cD}{{\cal D}}
\nc{\cE}{{\cal E}}
\nc{\cG}{{\cal G}}
\nc{\cH}{{\cal H}}
\nc{\cL}{{\cal L}}
\nc{\cO}{{\cal O}}
\nc{\cT}{{\cal T}}
\nc{\cN}{{\cal N}}
\nc{\rvac}[1]{|{\cal O}#1\rangle}
\nc{\lvac}[1]{\langle{\cal O}#1|}
\nc{\rvacb}[1]{|{\cal O}_\beta #1\rangle}
\nc{\lvacb}[1]{\langle{\cal O}_\beta #1 |}
\nc{\bb}{\bar{\beta}}
\nc{\bt}{\tilde{\beta}}
\nc{\ctH}{\tilde{\cal H}}
\nc{\chH}{\hat{\cal H}}
\nc{\1}{\aa}
\nc{\2}{\"{a}}
\nc{\3}{\"{o}}
\nc{\4}{\AA}
\nc{\5}{\"{A}}
\nc{\6}{\"{O}}
\nc{\al}{\alpha}
\nc{\g}{\gamma}
\nc{\Del}{\Delta}
\nc{\e}{\epsilon}
\nc{\eps}{\epsilon}
\nc{\lam}{\lambda}
\nc{\om}{\omega}
\nc{\Om}{\Omega}
\nc{\ve}{\varepsilon}
\nc{\mn}{{\mu\nu}}
\nc{\k}{\kappa}
\nc{\vp}{\varphi}

\nc{\advp}[3]{{\it  Adv.\ in\ Phys.\ }{{\bf #1} {(#2)} {#3}}}
\nc{\annp}[3]{{\it  Ann.\ Phys.\ (N.Y.)\ }{{\bf #1} {(#2)} {#3}}}
\nc{\apl}[3]{{\it  Appl. Phys. Lett. }{{\bf #1} {(#2)} {#3}}}
\nc{\apj}[3]{{\it  Ap.\ J.\ }{{\bf #1} {(#2)} {#3}}}
\nc{\apjl}[3]{{\it  Ap.\ J.\ Lett.\ }{{\bf #1} {(#2)} {#3}}}
\nc{\app}[3]{{\it Astropart.\ Phys.\ }{{\bf #1} {(#2)} {#3}}}
\nc{\cmp}[3]{{\it  Comm.\ Math.\ Phys.\ }{{ \bf #1} {(#2)} {#3}}}
\nc{\cqg}[3]{{\it  Class.\ Quant.\ Grav.\ }{{\bf #1} {(#2)} {#3}}}
\nc{\epl}[3]{{\it  Europhys.\ Lett.\ }{{\bf #1} {(#2)} {#3}}}
\nc{\ijmp}[3]{{\it Int.\ J.\ Mod.\ Phys.\ }{{\bf #1} {(#2)} {#3}}}
\nc{\ijtp}[3]{{\it Int.\ J.\ Theor.\ Phys.\ }{{\bf #1} {(#2)} {#3}}}
\nc{\jmp}[3]{{\it  J.\ Math.\ Phys.\ }{{ \bf #1} {(#2)} {#3}}}
\nc{\jpa}[3]{{\it  J.\ Phys.\ A\ }{{\bf #1} {(#2)} {#3}}}
\nc{\jpc}[3]{{\it  J.\ Phys.\ C\ }{{\bf #1} {(#2)} {#3}}}
\nc{\jap}[3]{{\it J.\ Appl.\ Phys.\ }{{\bf #1} {(#2)} {#3}}}
\nc{\jpsj}[3]{{\it J.\ Phys.\ Soc.\ Japan\ }{{\bf #1} {(#2)} {#3}}}
\nc{\lmp}[3]{{\it Lett.\ Math.\ Phys.\ }{{\bf #1} {(#2)} {#3}}}
\nc{\mpl}[3]{{\it  Mod.\ Phys.\ Lett.\ }{{\bf #1} {(#2)} {#3}}}
\nc{\ncim}[3]{{\it  Nuov.\ Cim.\ }{{\bf #1} {(#2)} {#3}}}
\nc{\np}[3]{{\it  Nucl.\ Phys.\ }{{\bf #1} {(#2)} {#3}}}
\nc{\pr}[3]{{\it Phys.\ Rev.\ }{{\bf #1} {(#2)} {#3}}}
\nc{\pra}[3]{{\it  Phys.\ Rev.\ A\ }{{\bf #1} {(#2)} {#3}}}
\nc{\prb}[3]{{\it  Phys.\ Rev.\ B\ }{{{\bf #1} {(#2)} {#3}}}}
\nc{\prc}[3]{{\it  Phys.\ Rev.\ C\ }{{\bf #1} {(#2)} {#3}}}
\nc{\prd}[3]{{\it  Phys.\ Rev.\ D\ }{{\bf #1} {(#2)} {#3}}}
\nc{\prl}[3]{{\it Phys.\ Rev.\ Lett.\ }{{\bf #1} {(#2)} {#3}}}
\nc{\pl}[3]{{\it  Phys.\ Lett.\ }{{\bf #1} {(#2)} {#3}}}
\nc{\prep}[3]{{\it Phys\. Rep.\ }{{\bf #1} {(#2)} {#3}}}
\nc{\prsl}[3]{{\it Proc.\ R.\ Soc.\ London\ }{{\bf #1} {(#2)} {#3}}}
\nc{\ptp}[3]{{\it  Prog.\ Theor.\ Phys.\ }{{\bf #1} {(#2)} {#3}}}
\nc{\ptps}[3]{{\it  Prog\ Theor.\ Phys.\ suppl.\ }{{\bf #1} {(#2)} {#3}}}
\nc{\physa}[3]{{\it  Physica\ A\ }{{\bf #1} {(#2)} {#3}}}
\nc{\physb}[3]{{\it  Physica\ B\ }{{\bf #1} {(#2)} {#3}}}
\nc{\phys}[3]{{\it Physica\ }{{\bf #1} {(#2)} {#3}}}
\nc{\rmp}[3]{{\it  Rev.\ Mod.\ Phys.\ }{{\bf #1} {(#2)} {#3}}}
\nc{\rpp}[3]{{\it Rep.\ Prog.\ Phys.\ }{{\bf #1} {(#2)} {#3}}}
\nc{\sjnp}[3]{{\it Sov.\ J.\ Nucl.\ Phys.\ }{{\bf #1} {(#2)} {#3}}}
\nc{\spjetp}[3]{{\it Sov.\ Phys.\ JETP\ }{{\bf #1} {(#2)} {#3}}}
\nc{\yf}[3]{{\it Yad.\ Fiz.\ }{{\bf #1} {(#2)} {#3}}}
\nc{\zetp}[3]{{\it Zh.\ Eksp.\ Teor.\ Fiz.\  }{{\bf #1}  {(#2)} {#3}}}
\nc{\zp}[3]{{\it Z.\ Phys.\ }{{\bf #1} {(#2)} {#3}}}
\nc{\ibid}[3]{{\sl ibid.\ }{{\bf #1} {#2} {#3}}}
\nc{\rf}[1]{(\ref{#1})}
\nc{\nn}{\nonumber \\*}
\nc{\bfB}{\bf{B}}
\nc{\bfv}{\bf{v}}
\nc{\bfx}{\bf{x}}
\nc{\bfy}{\bf{y}}
\nc{\vx}{\vec{x}}
\nc{\vy}{\vec{y}}
\nc{\oB}{\overline{B}}
\nc{\oI}{\overline{I}}
\nc{\oR}{\overline{R}}
\nc{\rar}{\rightarrow}
\nc{\ti}{\times}
\nc{\slsh}{\hskip-5pt/}
\nc{\sm}{Standard~Model~}
\nc{\MP}{M_{\rm Pl}}
\nc{\tp}{t_{\rm Pl}}
\nc{\ave}{\bar{E}}

\nc{\eff}{{\rm eff}}
\nc{\kk}{\vek{k}}
\nc{\pp}{{\rm p}}
\nc{\ga}{g_{a\gamma}}
\nc{\vv}{\\}
\nc{\eee}{{\bf E}}
\nc{\bbb}{{\bf B}}
\nc{\qcd}{T_{\rm QCD}}
\nc{\G}{\rm \ G}
\def\vec#1{{\bf #1}}
\begin{document}

{\title{\vskip-2truecm{\hfill {{\small OUTP-97-49-P \\
\hfill TURKU-FL/27-97}}\vskip 1truecm}
{\bf Baryogenesis and the Thermalization Rate of Stop}}


{\author{
{\sc Kari Enqvist$^{1}$}\\
{\sl\small Department of Physics, P.O. Box 9,
FIN-00014 University of Helsinki,
Finland} \\
{\sc Antonio Riotto$^{2}$}\\
{\sl\small Theoretical Physics Department, University of Oxford, 1 Keble Road,
OX1 3NP, Oxford, UK }
\\
and\\
{\sc Iiro Vilja$^{3}$}\\
{\sl\small Department of Physics, University of Turku,FIN-20014 Turku,  
Finland}
}}
\maketitle
\vspace{2cm}
\begin{abstract}
\noindent
We take the first steps towards the complete computation of the
thermalization rate
of the supersymmetric particles involved in electroweak
baryogenesis by computing 
the thermalization rate of the right-handed stop from 
the imainary part of the two-point Green function. We use improved propagators
including resummation of hard thermal loops. The thermalization rate
is computed at the one-loop level in the high temperature approximation as
a function of $M_{\tilde t_R}(T)$. We also give
an estimate for the magnitude of the two-loop contributions which dominate 
the rate for small $M_{\tilde t_R}(T)$. If the stop is non-relativistic
with $M_{\tilde t_R}(T)\gg T$, thermalization takes place by decay
and is very fast.

\end{abstract}
\vfil
\footnoterule
{\small $^1$enqvist@pcu.helsinki.fi};
{\small $^2$riotto@thphys.ox.ac.uk};
{\small $^3$vilja@newton.tfy.utu.fi.}
\thispagestyle{empty}
\newpage
\setcounter{page}{1}

It is now commonly accepted that the generation of the baryon  
asymmetry during the electroweak phase transition \cite{reviews} requires
some new physics at the weak scale. 
Threfore electroweak  
baryogenesis in the framework of the Minimal Supersymmetric Standard Model  
(MSSM) has  attracted much attention in the past years, with 
particular emphasis on the strength of the phase transition ~\cite{early} and  
the mechanism of baryon number generation \cite{nelson,noi,ck}.
 It has  recently been shown both analytically \cite{r1,color} and by lattice  
simulations \cite{r2}  that the phase transition can be sufficiently strongly  
first order if the lightest  stop is not much heavier than the top quark, the
ratio of the vacuum expectation values of the two neutral Higgses $\tan\beta$  
is smaller than $\sim 4$ and the lightest Higgs  is  lighter than about 85 
GeV. 

Moreover, the MSSM may contain, besides the CKM matrix phase, new  
CP-violating phases in the soft supersymmetry breaking  
parameters associated with the stop mixing angle and with  the gaugino and  
neutralino mass matrices. In the MSSM  large values of the stop mixing angle
are  restricted in order to preserve a
sufficiently strong first order electroweak phase transition.
Therefore, an acceptable baryon asymmetry
may only be generated through a delicate balance between the values
of the different soft supersymmetry breaking parameters contributing
to the stop mixing parameter, and their associated CP-violating
phases \cite{noi,mv}. 
In the MSSM
 it turns out that the main  
contribution to the  baryon asymmetry comes from
charginos and neutralinos and  the phase of the parameter $\mu$ must be    
larger than about 0.1 to be responsible for the generation of the observed  
baryon asymmetry \cite{noi,mv,ck}. If the strength of the
 electroweak phase transition is enhanced by the presence of some
new degrees of freedom beyond the ones contained in the MSSM, {\it
e.g.} some extra standard model gauge singlets,
 light stops (predominantly the 
right-handed ones) and charginos/neutralinos are expected to 
give quantitatively the
same contribution to the final baryon asymmetry.

Extra and sizeable  CP-violating phases in the MSSM are therefore a necessary  
ingredient for a successful electroweak baryogenesis scenario. 
If the particles  involved in the process of baryon number generation 
thermalize rapidly, CP-violating   
sources however 
loose their coherence and are diminished. This phenomenon can be  
intuitively understood by means of the following example: let us focus on the  
right-handed stop current and  imagine that a  right-handed stop scatters off  
the advancing bubble wall and is transformed into a left-handed stop. If the  
latter scatters off the Higgs background once again, a right-handed stop   
reappears in the plasma; since in both interactions with the wall CP is  
violated at the vertices because of the explicit phase in $A_t\mu$, quantum  
interference may give rise to   a final  right-handed current.  This, 
however,  
only takes place if,  along their way,  the stops do not interact with the  
surrounding plasma and disappear: large thermalization rates reduce   the  
final baryon asymmetry.


Moreover,  CP-violating currents in supersymmetric baryogenesis  are  
more easily built up if the degrees of freedom  in the stop  and in  the  
gaugino/neutralino 
sectors are nearly degenerate in mass \cite{noi}. For instance,  
phases of $\mu$  smaller than 0.1 are only consistent with
the observed baryon asymmetry for values of $|\mu|$ of the
order of the gaugino mass parameters. This is due to a large
enhancement of the computed baryon asymmetry for these values
of the parameters. This resonant behaviour is  
associated with the
possibility of absorption (or emission) of Higgs quanta by the
propagating supersymmetric particles. For momenta of the order of the critical
temperature, this can only take place when, for instance,  the Higgsinos and 
gauginos do not differ too much in mass.
By using the Uncertainty Principle, it is easy to understand that the
width of this resonance is expected
to be proportional to the thermalization rate  of the particles giving rise to
the baryon asymmetry \cite{noi}.

It is therefore clear that the computation of the thermalization 
rate of  
the particles responsible for supersymmetric baryogenesis represents a  
necessary step towards  the final computation of the baryon number. Despite    
its relevance,  no substantial effort has been devoted to a detailed  
computation of the decay width of SUSY particles in the thermal bath. 
The goal of this paper is to take the first step 
towards a complete evaluation of the thermalization rate of the  supersymmetric
particles involved in the 
generation of the baryon asymmetry, right-handed stops, charginos and
neutralinos. At present, we restrict ourselves to the
 computation of  the thermalization rate  
$\Gamma_R$  
of light  
right-handed stops from the imaginary part of the two-point Green function in  
the {\it unbroken} phase of the MSSM. We use improved propagators including   
resummation of hard thermal loop;  the thermalization rate is computed at the
one-loop level in the high temperature approximation
and an estimate is given for two-loop contributions. 
Finite mass effects 
as well as the computation of the thermalization rate for charginos and
neutralinos will be presented in a longer publication \cite{inprep}. 

Let us first   present some technical details needed for the  
computation of the right-handed stop  
thermalization rate. A great deal of the formalism may be found in    
\cite{EEV}, but 
we briefly summarise here some details  and also present  some new necessary  
technical tools.

The thermalization rate $\Gamma_R$ depends upon the imaginary part of the two  
point Green function $\Sigma_R$ via the relation 
\be{gammadef}
\Gamma_R=-{{\rm  
Im}\:\Sigma_R\over \omega}~, 
\ee
where $\omega=\omega(k)$ is the energy of a given mode. In the present paper
we shall focus on the long wave-length stops with $\omega\simeq 
M_{{\tilde t}_R}$.
At the one-loop level the quantity  ${\rm Im}\:\Sigma_R$ receives   
different   
contributions  from  diagrams involving   SM fermions, charginos,  
neutralinos,  scalar and gauge bosons\footnote{We assume R-parity 
conservation so that  one of the  
two particles running in the loop is always a superpartner.}. 
At two loop level there are several potentially relevant sources that 
contribute to the thermalization rate. The strong interactions contribute,
due to diagrams including gluon interactions being proportional to the 
square of the strong coupling constant. Also, there are possible 
interactions of two right-handed stops appearing from $D$-terms of the 
superpotential. They are, however, proportional to the fourth power of the
electroweak gauge couplings (and so are two-loop diagrams containing 
electroweak gauge bosons) and therefore can be neglected. Possible 
interactions arise also from the $F$-terms 
of the superpotential. Of particular importance
is the one proportional to the top-quark Yukawa coupling given by
$h_t^2 |H_2|^2 |\tilde t_R|^2$. The diagram including (only) this interaction 
is not suppressed by the  coupling, but is expected to be small due to 
phase space suppression unless the right-handed stop is light. The very 
same argument can be used in the case of gluon interactions, too.

The imaginary parts are cuts across the relevant  diagrams and 
correspond to the differences between the absorption and the emission
rates, or between decay and inverse decay rates. For brevity 
we denote these respectively
by ``absorption''  and ``decay''. For all the fermions and bosons 
unbroken phase thermal corrections must be taken into account. They merely  
change the pole structure of the propagators and, to leading order in the  
temperature $T$ and for the massless SM fermions, they are the hard thermal  
loops. Accordingly, there appear energy thresholds which,  in particular for  
the fermionic loops,  are rather complicated because of the 
complicated nature  
of fermionic dispersion relations, which have two branches called particles and
holes \cite{weldon}. However, the vertex corrections relevant in the present 
discussion 
contains no hard thermal loops and can therefore be omitted \cite{BP&HMSV}.

The plasma corrected left-handed Dirac
fermion propagator reads generically ($P_\mu = ( p_0,\, {\bf p})$, and
$p = |{\bf p}|$) 

\be{propag}
S(P) = P_L {F_\mu \gamma^\mu + \mu\over F^2 - \mu^2}P_R,
\ee
where $P_L$ and $P_R$ are the left- and right-handed projections and
$\mu$ is the bare mass, and
\be{proj}
\gamma_\mu F^\mu = [1 + a(P)] P_\mu\gamma^\mu + b(P)\gamma^0.
\ee
In the case of a left-handed field,
at high temperature $T\gg\mu$ the functions $a$ and $b$ are given by  
\cite{weldon}
\be{hight}
a_L(P) = {m^2_L\over p^2} \left ( 1 - {p_0\over 2p} \ln \left | {p_0 + p\over
p_0 - p}\right | \right )
\ee
and
\be{b(p)}
b_L(P) = {m^2_L\over p^2} \left [ - {p_0\over p} + \frac 12 \left(
{p_0^2\over p^2} - 1 \right ) \ln \left | {p_0 + p\over
p_0 - p}\right | \right ],
\ee
where $m_L$ is the plasma mass of the left-handed particle.
 An identical formula 
can be written for right-handed fermion component with $m_L$ replaced by
$m_R$. 

\eqs{hight}{b(p)} apply to all particles with masses $\ll T$, that is, to
all the SM fermions in the unbroken  
phase. 
For Dirac fermions with non-zero bare mass $m$, the situation is more 
complicated because the left- and right-handed  components couple. 
 The Majorana propagators for massive neutralinos would be even
more complicated \cite{RiottoV}. In the present paper we shall 
focus on the limit $T\gg m$, which we believe 
cover most of the cases of interest. 
In present context the only 
Dirac particles with non-zero bare mass are the top quarks and the
charginos, whose  
right- and left-handed plasma
masses are equal. The bare mass of  the 
charged gauginos $\widetilde{W}^{\pm}$ is  
the $SU(2)_L$ soft supersymmetry breaking gaugino mass $M_2$ and the bare 
mass  
$\mu$ of charged Higgsinos $\widetilde{H}^{\pm}$  emerges from the bilinear  
term $\mu \hat{H}_1\hat{H}_2$ in the superpotential. 

We assume that the 
only particles  light enough to be in equilibrium with the thermal bath at  
temperature $T$ are the SM
particles, the right-handed stop $\widetilde{t}_R$, the charginos  
$\widetilde{W}^{\pm}$ and $\widetilde{H}^{\pm}$ , the neutralinos  
$\widetilde{B}$, $\widetilde{W}_3$ and $\widetilde{H}^{0}_{1,2}$, and the  
neutral and charged scalar fields in the two  Higgs doublets. 
This amounts to assuming  that   all the other supersymmetric  particles,  
{\it e.g.} left-handed stops and gluinos, are much heavier than  $T$. 
This choice is motivated by considerations about 
the  strength of the phase transition \cite{r1,color,r2} and the mechanism for 
baryon  
number generation \cite{nelson,noi,ck} in the MSSM. For instance, 
 a strongly first order electroweak phase transition
can  be achieved in the presence of a top squark
not much heavier than the top quark \cite{r1,color}.
In order to naturally
suppress its contribution to the parameter $\Delta\rho$ and hence
preserve a good agreement with the precision measurements at LEP,
it should be mainly right-handed. This can be achieved if the 
soft supersymmetry breaking mass $m_Q$ of $\widetilde{t}_L$
is much larger than $M_Z$ so that $\widetilde{t}_L$ is decoupled from the  
thermal bath. It is important to keep in mind that this mass  hierarchy
between the left- and right-handed stops may be relaxed if the strength
of the phase transition is enhanced by light boson degrees of freedom
other than the right-handed stops themselves. This happens, for instance, if
the MSSM content is increased by adding a standard model gauge singlet.  

The relevant hight $T$ plasma  
masses are given in the Table 1. We have separated the SM   
contribution $m_{{\rm SM}}$ from the  MSSM contribution $m_{{\rm MSSM}}$, so  
that the plasma mass is actually  the sum of the two.
We also  denote by $g_1=0.247,\ g_2=0.640$ and $ g_3=1.243$ respectively
the values of $U(1)_Y$, 
$SU(2)_L$ and $SU(3)_c$ couplings at $T\simeq M_W$,
and we take the (top-quark) Yukawa coupling to be $h_t=1$. 
Other couplings and
interactions are not included because of their relative smallness.
The relevant fermionic one-loop diagrams are those, where the right-handed stop 
couples to one of the pairs $\tilde H_2^0\, t_L$, $\tilde H_2^+\, b_L$ or
$\tilde B\, t_R$. The possible bosonic one-loop diagrams consist of 
those including $\tilde t_R$ and U(1) gauge boson $B$ or $\tilde t_R$ and gluon.


\begin{table}
\centering
\begin{tabular}{lcc}
Particle& $m_{{\rm SM}}^2/T^2$& $m_{{\rm MSSM}}^2/T^2$\\ \hline

$W_{1, 2, 3}$	& $\frac 29 g_2^2$ 	& $\frac {11}{18} g_2^2$ \\ 

$B$		& $\frac {21}{36} g_1^2$& $\frac {21}{27} g_1^2$ \\ 

$g$		& $\frac 23 g_3^2$	& $\frac 1{18} g_3^2$ \\ 

$t_L,\ b_L$	& $\frac 16 g_3^2 + \frac 3{32} g_2^2 + \frac 1{216} g_1^2 +
\frac 1{16} h_t^2$	& $\frac 1{16} h_t^2$ \\ 

$t_R$		& $\frac 16 g_3^2 + \frac 1{18} g_1^2 + \frac 18 h_t^2$
& $\frac 4{54} g_1^2$ \\ 

$b_R$		& $\frac 16 g_3^2 +
\frac{1}{72} g_1^2+\frac 1{16} h_t^2$			& 0 \\ 

$H_1^0,\ H_1^{-}$
	& $\frac{3}{16} g_2^2+\frac{1}{16}g_1^2$			 
& $\frac{3}{16} g_2^2 + \frac {3}{48} g_1^2$\\ 

$H_2^0,\ H_2^{+}$	&  $\frac{3}{16} g_2^2+\frac{1}{16}g_1^2	 
+\frac{1}{4}h_t^2$		&  $\frac{3}{16} g_2^2 + \frac {3}{48} g_1^2  
+\frac{1}{4}h_t^2$\\ 

$\widetilde t_R$	& 0			& $\frac 49 g_3^2 + \frac 13  
h_t^2 +
\frac{36}{108} g_1^2$\\ 

$\widetilde H_1^0,\ \widetilde H_1^{-}$ & 0	&$\frac 3{32} g_2^2 + \frac 18  
g_1^2$\\ 

$\widetilde H_2^0,\ \widetilde H_2^{+}$ & 0	& $\frac 1{16} h_t^2 + \frac  
3{32} 

g_2^2 + \frac 18 g_1^2$\\ 

$\widetilde B$	& 0			& $\frac 29 g_1^2$ \\ 

$\widetilde W_{1,2,3}$& 0			& $\frac 3{16} g_2^2$ \\ 

\end{tabular}
\caption{Plasma masses of light particles contributing to stop decay rate. }
\end{table}

Let us first consider the fermionic contribution to the 
one-loop imaginary part of the right-handed stop
self-energy. 
The 
right-handed stop can  absorb a right-handed Higgsino hole 
(left-handed quark hole)
to produce a left-handed quark particle (right-handed Higgsino particle),
or it can decay into a pair of particles or a pair of holes.
The Higgsino may be either a neutralino or a chargino.
In the  high temperature
approximation, where the bare masses of the particles 
running in the loop can always be neglected but hard thermal
corrections to the propagators are included, the
absorption contribution to ${\rm Im}\:\Sigma_R$ is given by 
\cite{EEV}
\bea {abs}
{\rm Im}\:\Sigma_{abs}&=&{4e_f^2\over \pi}{k^2(\omega_{L,p}^2-k^2)
(\omega_{R,h}^2-k^2)\over 16m_L^2m_R^2}\left[n^+(\beta\omega_{R,h})
-n^+(\beta\omega_{L,p})\right]
\\ \nonumber
& + &(R\leftrightarrow L,~
h\leftrightarrow p)~.
\eea
Here $(R,h)$ and $(L,p)$ refer to right-handed holes and left-handed particles,
respectively, and $\omega_{L,p}(k)$ and $\omega_{R,h}$ are solutions to the
fermion dispersion relations, given by 
\bea{dispers}
\omega_{p,h} &=& - a(\omega_{p,h}, k) \omega_{p,h} - b(\omega_{p,h}, k) 
\pm (1 + a(\omega_{p,h}, k)) 
k\\ \nonumber
&=& \pm \left [ k + {m^2\over k}\left[1+\frac12\left(\pm 1-{\omega_{p,h}
\over k}\right)
{\rm ln}\left({\omega_{p,h}+k\over \omega_{p,h}-k}\right)\right]\right ]~,
\eea
where the upper and lower signs refer to particles and holes, respectively, and
$m$ is the appropriate plasma mass. The solutions are
depicted in Fig. \ref{kuva2} for the
case of $t$ and $\tilde H$. The coupling factor $e_f^2=h_t^2$ for the loops
involving $t_L$ and $b_L$ (at high $T$ their contributions are equal),
and $e_f^2=g_1^2/2$ for the case of $\tilde B t_R$-loop.
In \eq{abs} we have defined
\be{disp}
n^{\pm}(x)\equiv {1\over e^{x}\pm 1}~.
\ee
Taking $\tilde t_R$ to be at rest, the 
momentum $k$ is determined by the energy conservation condition
\be{cons}
M_{{\tilde t}_R}+\omega_{R,h}=
\omega_{L,p}~~~~(M_{{\tilde t}_R}+\omega_{L,h}=\omega_{R,p})~. 
\ee
\begin{figure}
\leavevmode
\centering
\vspace*{50mm} 
\includegraphics{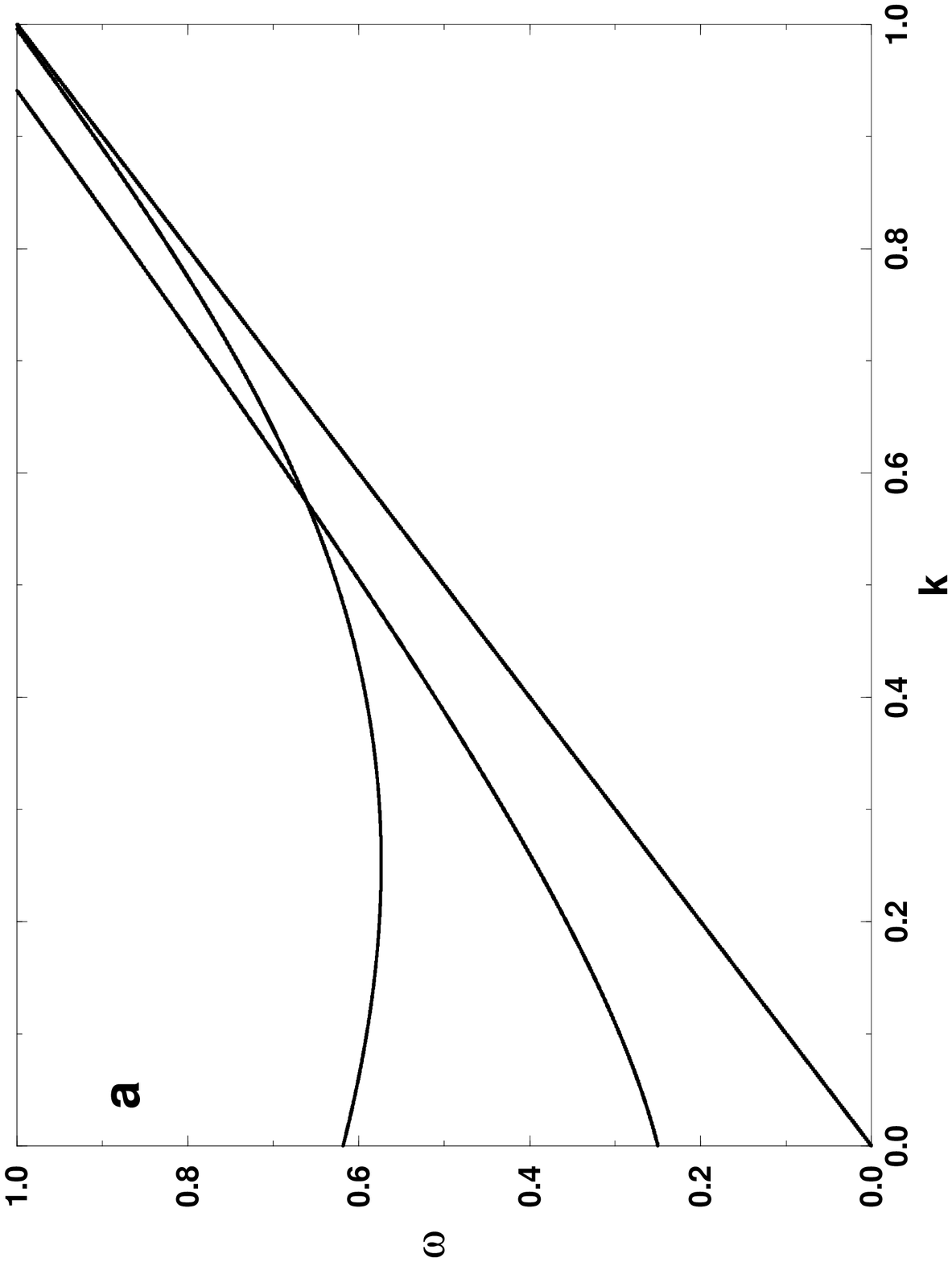}
\includegraphics{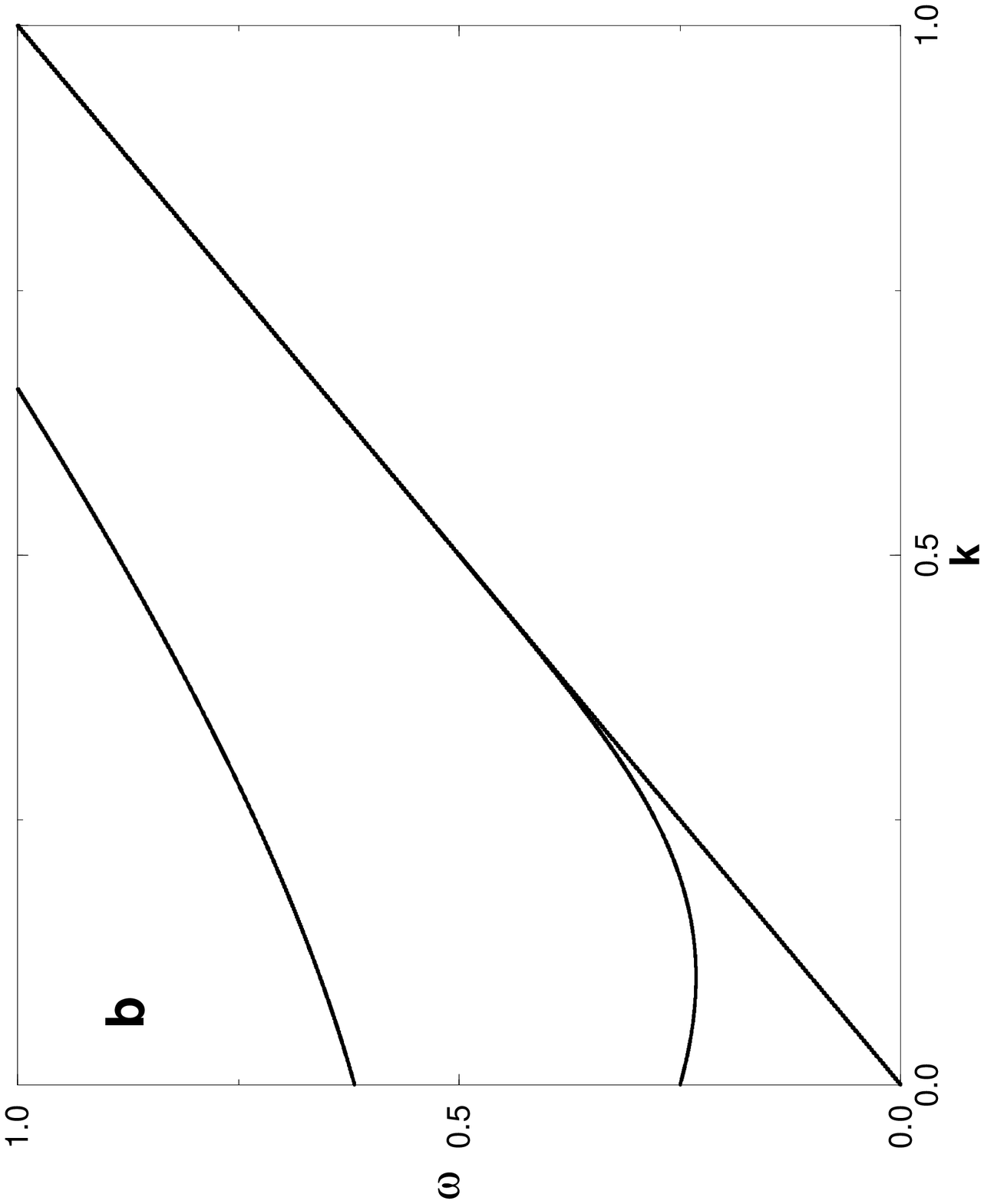}
\caption{Fermionic dispersion relations: (a) $t$-hole, $\tilde H_2$-particle;
(b) $t$-particle, $\tilde H_2$-hole.}
\label{kuva2}       
\end{figure} 
At high $T$ the direct decay channel contributes an imaginary part \cite{EEV}
\bea{decay}
{\rm Im}\:\Sigma_{dec}&=&{4e_f^2\over \pi}{k^2(\omega_{L,p}^2-k^2)
(\omega_{R,p}^2-k^2)\over 16m_L^2m_R^2}\left[1-n^+(\beta\omega_{R,p})
-n^+(\beta\omega_{L,p})\right]
\\ \nonumber
& + &(p\leftrightarrow h)~,
\eea
where $k$ is determined by 
\be{energydec}
M_{{\tilde t}_R}=\omega_{R,a}+\omega_{L,a}
\ee
with
$a=p,h$. At high temperatures the plasma mass of $\tilde t_R$ is 
$M_{{\tilde t}_R}(T)=1.020T$, whereas $M_{\tilde H_{2}}(T)=0.329T$,
$M_{\tilde B}(T)=0.116T$, 
$M_{b_{L}}(T)=M_{t_{L}}(T)=0.649T$, and $M_{t_{R}}(T)=0.625T$
so that decay is clearly kinematically possible
to both $\tilde B t_R$ and $\tilde H_{2}t_L$ or $\tilde H_{2}b_L$.
The full fermionic high $T$
contribution to ${\rm Im}\:\Sigma_R$ is thus given by the sum
$2\:{\rm Im}\:\Sigma_{abs}(\tilde H_{2}t_L)+
{\rm Im}\:\Sigma_{abs}(\tilde B t_R)
+\:2\:{\rm Im}\:\Sigma_{decay}(\tilde H_{2}t_L)+
{\rm Im}\:\Sigma_{decay}(\tilde B t_R)$, and it can only be
computed numerically by solving $k$ from \eqs{cons}{energydec} for a 
fixed $M_{{\tilde t}_R}(T)$, using the fermionic dispersion relations
for holes and particles as given by \eq{dispers}. (The factor 2 comes
about because the equality of $\tilde H_{2}t_L$ and $\tilde H_{2}b_L$
contributions.)

The high $T$ fermionic contributions to $\Gamma_R$ are depicted in Fig. 
\ref{kuva3}, plotted against 
$M_{{\tilde t}_R}(T)$, which is related to the zero temperature stop mass by
$M_{{\tilde t}_R}^2(T)=M_{{\tilde t}_R}^2(0)+\left(\frac{4}{9}g_3^2+
\frac{1}{3}h_t^2 +\frac{36}{108}g_1^2\right)T^2$. Note that 
for the fermionic contributions the high $T$
approximation does not concern $\tilde t_R$; the results apply equally
to the case where $\tilde t_R$ is no longer relativistic 
$(T\ll M_{{\tilde t}_R})$ as well as close to the phase transition, where
cancellation between the zero temperature and high temperature mass
terms is possible so that $T\gg M_{{\tilde t}_R}$.
Thus for a fixed
stop mass, from the figures one can read 
the various contributions to $\Gamma_{R}$ at a fixed temperature.

\begin{figure}
\leavevmode
\centering
\vspace*{50mm} 
\includegraphics{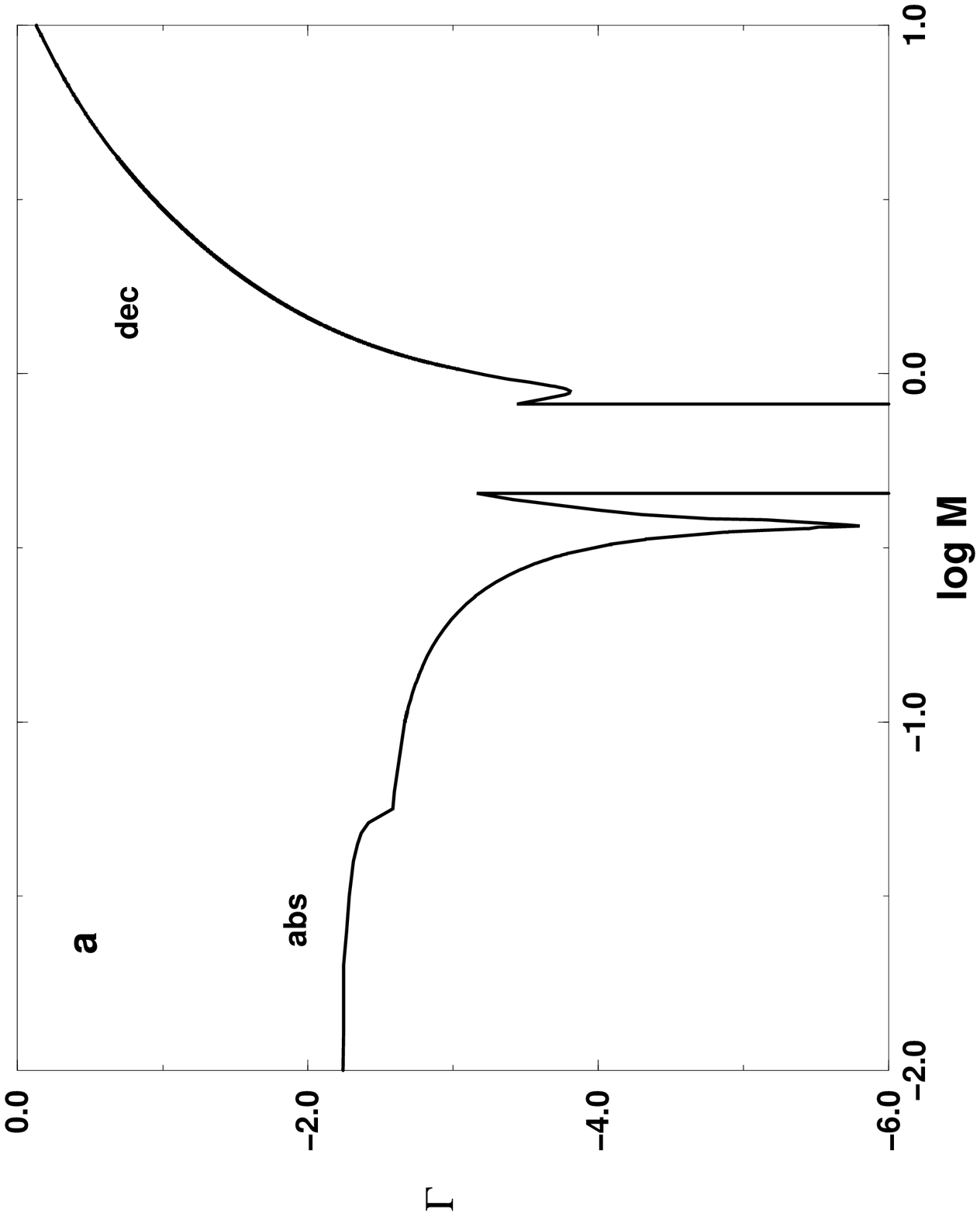}
\includegraphics{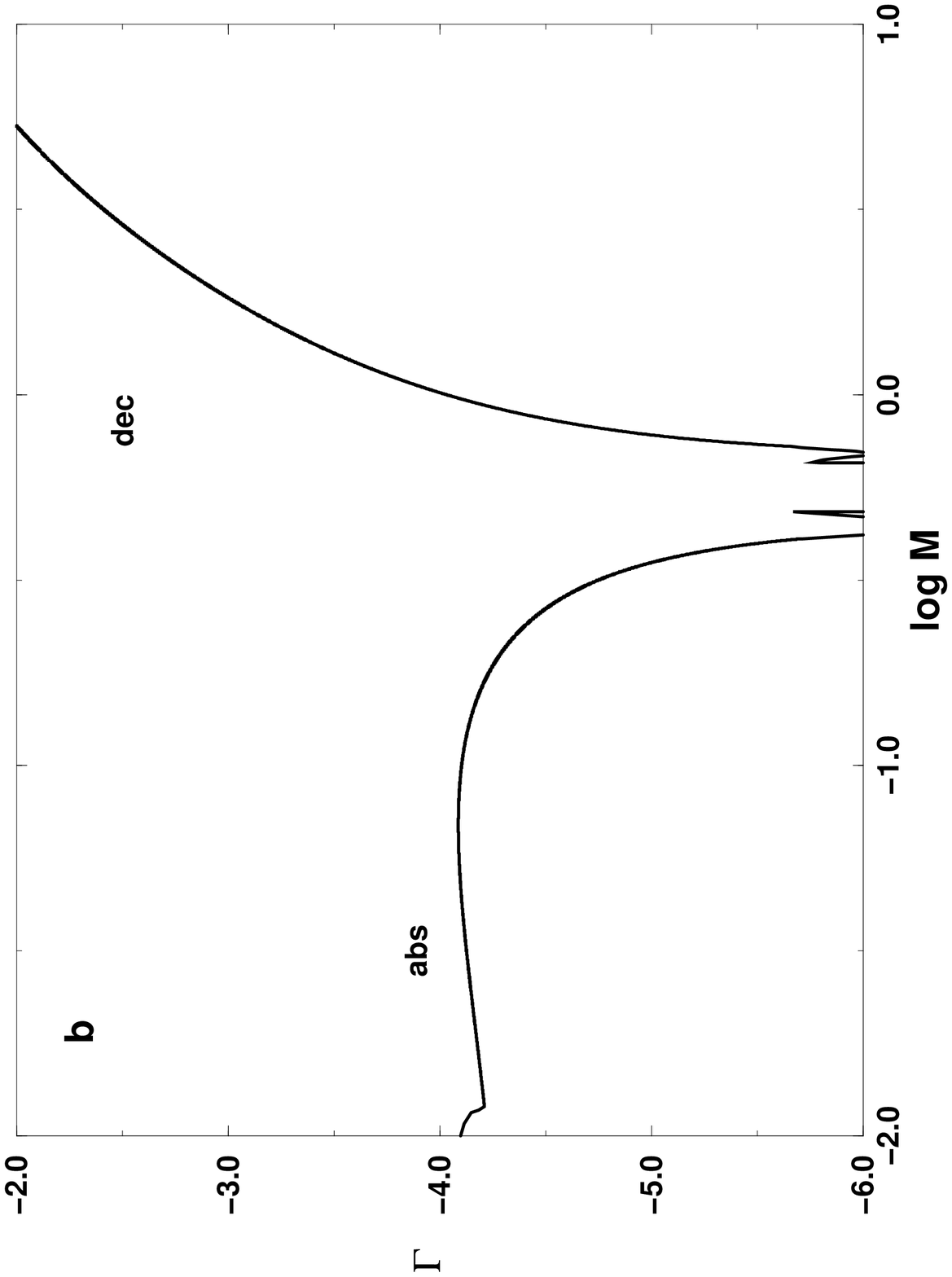}
\caption{Fermionic contributions to $\Gamma_R$ at high $T$: 
(a) $\tilde B$-$t_R$ -contribution; (b) $\tilde H$-$t_L$ -contribution. 
The curves refer to absorption and decay channels as explained in the text.
Both the 
decay rate and the mass are given in the units of temperature in 
a logarithmic scale.}
\label{kuva3}       
\end{figure} 

The one-loop, high $T$ gauge boson contribution to ${\rm Im}\:\Sigma_R$
can be cast in the form \cite{EEV}
\be{gauge}
{\rm Im}\:\Sigma_{gb}=-{e^2\over \pi}{\omega_g M_{{\tilde t}_R}^2k^4
\over \Pi (M_g^2-\Pi)\omega_{\tilde t}}
\left[n^-(\beta\omega_{\tilde t})-n^-(\beta\omega_{g})\right]~,
\ee
where  
$\Pi(\omega_g,k)$ is the longitudinal gauge boson self energy, given by
\be{piL}
\Pi(\omega_g,k)=3M_g^2\left(1-{\omega_g^2\over k^2}\right)
\left[1-\frac{\omega_g}{2k}
{\rm ln}\Big\vert{\omega_g+k\over\omega_g-k}\Big\vert\right]~.
\ee
$\omega_g$ is solved implicitly through
$\omega_g^2=k^2+\Pi$, and $\omega_{\tilde t}^2=k^2+M_{{\tilde t}_R}^2$.
The coupling factor $e^2$ are  $4 g_3^2/3$ and $4 g_1^2/9$
for gluons and $B$, respectively.
The energy conservation condition in this case is
\be{generg}
M_{{\tilde t}_R}+\sqrt{k^2+M_{{\tilde t}_R}^2}=\omega_g~. 
\ee
Physically \eq{gauge} corresponds to absorption; for kinematic reasons 
decay is not possible for gauge loops. 
Here we have implicitly assumed that $T\gg
|M_{{\tilde t}_R}(0)|$ so that $M_{{\tilde t}_R}(T)\le 1.02T$. (If we assume
$M^2_{{\tilde t}_R}(0)$ to be negative, it is possible that 
$M_{{\tilde t}_R}(T)\ll T$.) 
This is necessary because of the implementation of the high $T$ approximation 
in the loop, which now also concerns ${\tilde t}_R$ as it is circulating
in the loop. Hence the gauge boson
contribution \eq{gauge}
is not correct for non-relativistic ${\tilde t}_R$, but it
is nevertheless valid even 
if there is a cancellation between the stop bare mass and plasma mass terms.

The gauge boson contributions to $\Gamma_R$ at high $T$,
plotted against 
$M_{{\tilde t}_R}(T)$, are shown in Fig. \ref{kuva4}. One sees that 
gauge bosons do not contribute to $\Gamma_R$
when $\tilde t_R$ is in full equilibrium, but may be important 
if $M_{{\tilde t}_R}(T)\lsim T$. 
\begin{figure}
\leavevmode
\centering
\vspace*{60mm} 
\includegraphics{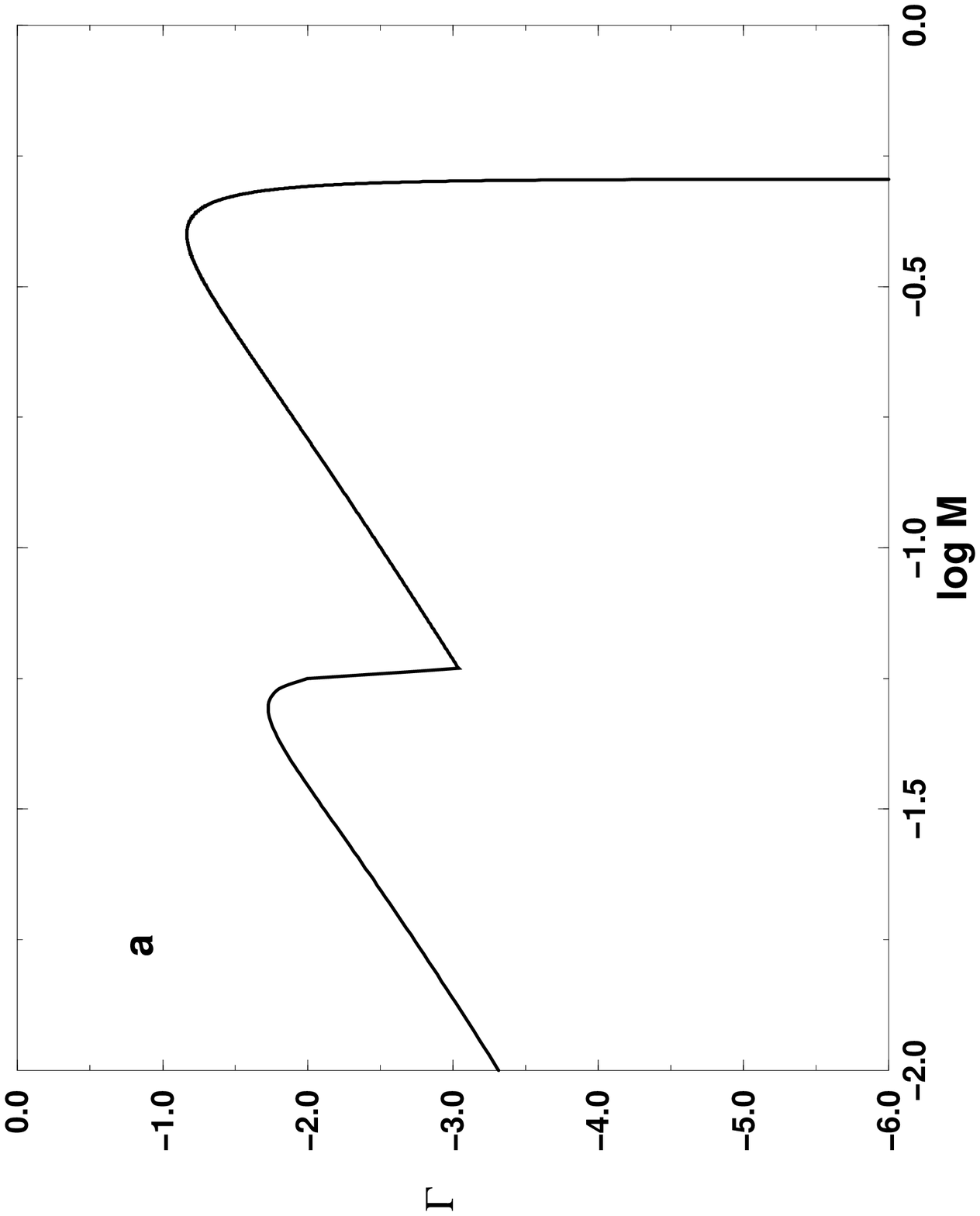}
\includegraphics{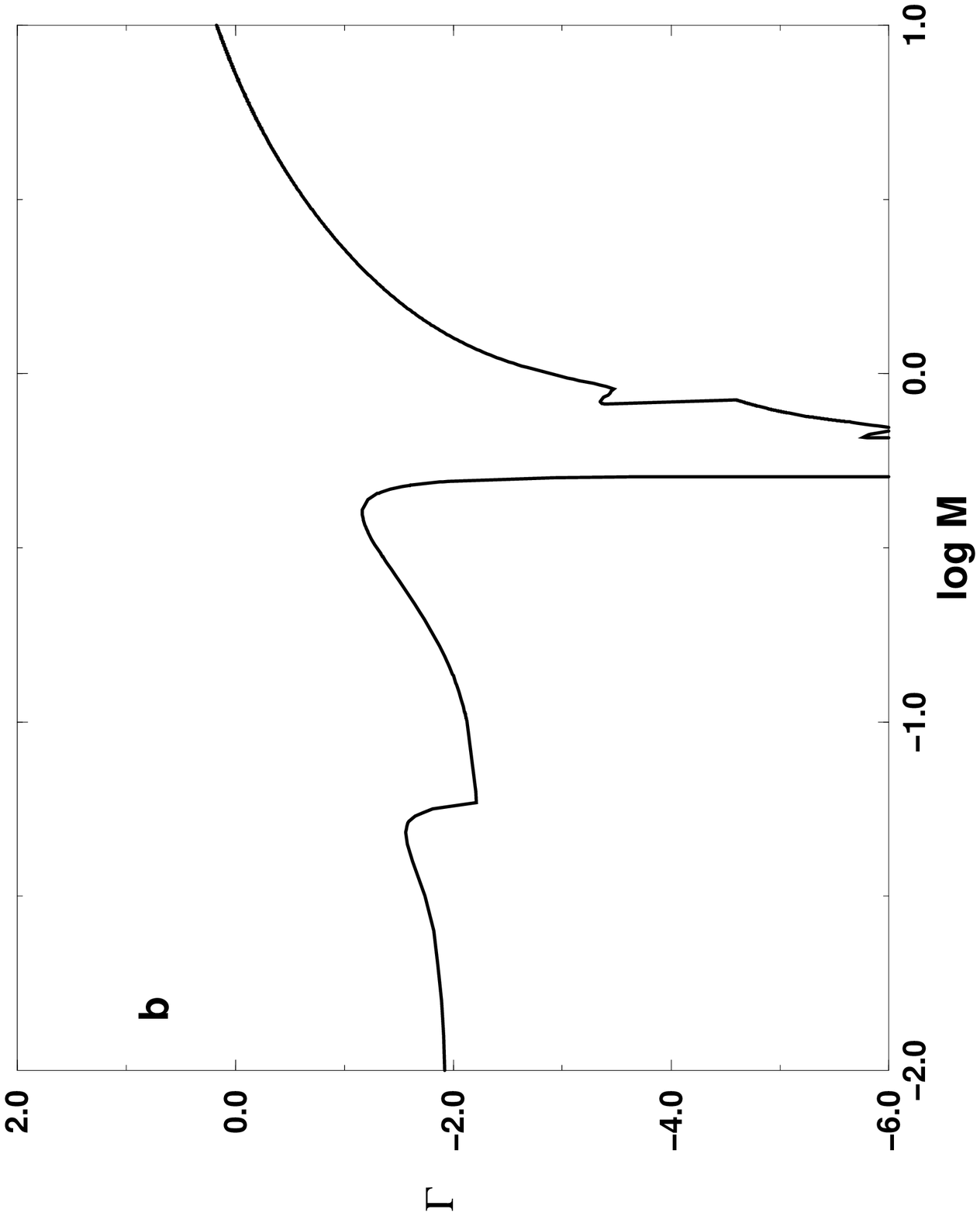}
\caption{(a) The gauge boson contributions to $\Gamma_R$ at high $T$; (b)
the total $\Gamma_R$. Both the decay rate and the mass are given in the units 
of temperature in a logarithmic scale. }
\label{kuva4}       
\end{figure} 
 In Fig. \ref{kuva4} we also
show the total 
high $T$ thermalization rate $\Gamma_{R}$.
At high temperature, far from the the criticial point, one finds that
at one loop
$\Gamma_{R}\simeq 8\times 10^{-4}T$.

The total one-loop result presented in Fig. \ref{kuva4} has a range in which 
$\Gamma_R$ is vanishingly small. This 
reflects only the kinematics of absorption and decay and will disappear
at higher loops. For instance, at two loops the cuts through various
diagrams give rise in addition to absorption and decay (involving more
than two particles in the initial or final state) to processes which
correspond to ordinary scattering. An example of the two-loop 
contributions are the so-called sunrise-diagrams appearing both from
gluon coupling and quartic Yukawa coupling (the electroweak gauge boson
can be neglected as discussed earlier). Neglecting the mass differences between
different loop particles, the imaginary parts can be estimated as in 
\cite{EEV}. We obtain $\Gamma_{R, h_t} \simeq h_t^4 (T^2/
128 \pi M_{{\tilde t}_R}(T))  = 0.0024 T^2/(M_{{\tilde t}_R}(T))$ and 
$\Gamma_{R, gluon} \simeq 7 g_3^4 T^2/(576 \pi M_{{\tilde t}_R}(T))
 = 0.0092 (T^2/M_{{\tilde t}_R}(T))$. There appear also
t-channel processes from gluon interactions. These may be also 
important whenever the right-handed stop is light. By making a rough 
approximation of the 
t-channel gluon processes in the spirit of ref. \cite{EEV},
we conclude that the t-channel contribution is comparable to the 
sunrise-diagrams; we estimate that
$\Gamma_{R,scatt}\simeq 3\times 10^{-3}T$ when 
$M_R(T) \sim T$. Thus two-loop diagrams are expected
to dominate the thermalization rate when $ 
M_{{\tilde t}_R}(T) \lsim 0.1T$, or in the range $0.5T\lsim
M_{{\tilde t}_R}(T) \lsim T$,
with $\Gamma_R\simeq 10^{-3}T^2/M_{{\tilde t}_R}(T)$.
In the range $0.1T\lsim
M_{{\tilde t}_R}(T) \lsim 0.5T$ absorption of gauge bosons dominates
the thermalization rate, and it can be as high as $0.1T$.

Note that right-handed stop plasma masses with 
$M_{{\tilde t}_R}(T)\ll T$ can occur only if $M^2_{{\tilde t}_R}(0)$
 is negative. Even though present experimental bounds on the stop masses
do not exclude such a possibility, one should keep in 
mind that $|M_{{\tilde t}_R}(0)|$ is bounded from above from considerations
about color breaking at zero temperature \cite{color}. If we restrict ourselves
to electroweak baryogenesis and therefore assume that
the critical temperature is $T_c={\cal O}(100)$ GeV, it is easy to
show that $M_{{\tilde t}_R}(T_c)$ must be larger than about $0.55\: T_c$
for masses of the lightest Higgs boson around $80$ GeV. (For lighter
Higgs boson the bound for $M_R$ is even larger.) 
                                                                       
Let us briefly discuss the implications of our findings. As we 
already 
mentioned in the introduction, the precise  knowledge of the 
thermalization rate
of the supersymmetric particles is a key ingredient for the 
computation of  the final
baryon asymmetry. Sizeable decay rates of the  particles 
propagating in the plasma
destroy  the quantum interference out of which the   the CP-violating sources
are built up  and therefore reduce the baryon asymmetry. Small decay rates,
on the other side, 
are   relevant when the particles reflecting off the advancing bubble 
wall  have
comparable masses and resonance effects show up  \cite{noi}. 
In such a case, the thermalization  rates provide  the natural width of these 
resonances and as the present calculation demonstrates, in supersymmetric
theories these depend in a complicated way on the particles involved
and their plasma masses.

Even though it is presently believed that the right-handed stops
do not play a leading role 
in generating the baryon asymmetry, 
it is important to emphasize that
this is only true   in  the context  of
the MSSM. There the phase transition is made strong by the infrared
effects in the right-handed stop sector 
and the baryon number is
 mainly generated by charginos and
neutralinos. Should the source of the strength of the phase transition
reside somewhere else, {\it i.e.} in Standard Model gauge singlets, 
the role of right-handed stops will  be comparable to the 
one played by charginos and higgsinos and the knowledge of the thermalization
rate $\Gamma_R$ is important to obtain  a precise estimate of the final baryon 
asymmetry.

If the stop is non-relativistic
with $M_{{\tilde t}_R}(T)\gsim T$, thermalization is dictated by the
one-loop
thermal decay rate which can be larger than $T$.
Thus in any case the thermalization of $\tilde t_R$
is rather fast, as can be seen in 
Fig. \ref{kuva4}.
Since the zero temperature limit on 
$|M_{{\tilde t}_R}(0)|$ from color breaking \cite{color} suggests that
$M_{{\tilde t}_R}(T)\gsim 0.55\:T$, we may conclude that during baryogenesis 
the thermalization rate of 
a relativistic
right-handed stop is dominated by two-loop effects (i.e.
scattering) with $\Gamma_R\simeq 10^{-3}T$. This is so in particular
because the absorption channels close at $M_{{\tilde t}_R}(T) 
\sim 0.51\: T$
before decay channels open at $M_{{\tilde t}_R}(T) \sim 0.66\: T$.
This means that the processes of quantum intereference, necessary 
build up the axial stop number, may be damped by the incoherent nature of the 
plasma if $\tilde t_R$ is non-relativistic at baryogenesis. Even in
the favorite case in which the left-handed and right-handed stops
have comparable masses, the resonance effects will be washed out by
the large right-handed stop thermalization rate. 

We hope to present our results about the thermalization rate of
charginos and neutralinos soon \cite{inprep}. 

\vskip 1cm
\subsection*{Acknowledgements}
We thank Per Elmfors for many useful discussions. KE is supported by 
the Academy of Finland.

\def\NPB#1#2#3{{\it Nucl. Phys.} {\bf B#1} (19#2) #3}
\def\PLB#1#2#3{{\it Phys. Lett.} {\bf B#1} (19#2) #3}
\def\PLBold#1#2#3{{\it Phys. Lett.} {\bf#1B} (19#2) #3}
\def\PRD#1#2#3{{\it Phys. Rev.} {\bf D#1} (19#2) #3 }
\def\PRL#1#2#3{{\it Phys. Rev.} Lett. {\bf#1} (19#2) #3}
\def\PRT#1#2#3{{\it Phys. Rep.} {\bf#1} (19#2) #3}
\def\ARAA#1#2#3{{\it Ann. Rev. Astron. Astrophys.} {\bf#1} (19#2) #3}
\def\ARNP#1#2#3{{\it Ann. Rev. Nucl. Part. Sci.} {\bf#1} (19#2) #3}
\def\MPL#1#2#3{{\it Mod. Phys. Lett.} {\bf #1} (19#2) #3}
\def\ZPC#1#2#3{{\it Zeit. f\"ur Physik} {\bf C#1} (19#2) #3}
\def\APJ#1#2#3{{\it Ap. J.} {\bf #1} (19#2) #3}
\def\AP#1#2#3{{\it Ann. Phys. } {\bf #1} (19#2) #3}
\def\RMP#1#2#3{{\it Rev. Mod. Phys. } {\bf #1} (19#2) #3}
\def\CMP#1#2#3{{\it Comm. Math. Phys. } {\bf #1} (19#2) #3}

\end{document}